# *In situ* synchrotron X-ray diffraction study of flash austenitization and process design insights in medium-Manganese steels for energy applications

*Bowen Zou* [a*], *Mathias Zapf* [b], *Thea Kannenberg* [c,d], *Daniel Schneider* [c,d], *Yixu Wang* [a], *Xiao Shen* [a], *Ulrich Prahl* [b], *Wenwen Song* [a]

[a]: Department of Granularity of Structural Information in Materials Engineering, Institute of Materials Engineering, University of Kassel, Mönchebergstraße 7, 34125 Kassel, Germany

[b]: Institute of Metal Forming, Faculty of Materials Science and Technology, Technische Universität Bergakademie Freiberg, Bernhard-von-Cotta-Straße 4, 09599 Freiberg, Germany

[c]: Institute of Digital Materials Science (IDM), Karlsruhe University of Applied Sciences (HKA), Moltkestraße 30, Karlsruhe, 76133, Germany

[d]: Institute for Applied Materials (IAM-MMS), Karlsruhe Institute of Technology (KIT), Straße am Forum 7, Karlsruhe, 76131, Germany

[*] Corresponding author: bowen.zou@uni-kassel.de

## Abstract

Medium Mn steels (MMnSs) are promising candidates for energy-related infrastructure because their multiphase microstructures and austenite stability can be tailored to improve failure resistance under demanding service conditions. Flash austenitization (FA) provides a rapid route to form austenite while limiting prior austenite grain coarsening and substitutional solute homogenization, but the related short-time transformation kinetics remain insufficiently quantified. In the present work, the effects of FA temperature and initial microstructure on austenitization kinetics were investigated in an Fe-6Mn-1.5Si-1Cr-0.3Mo-0.05Nb-0.2C (wt.%) MMnS using dilatometry-integrated *in situ* synchrotron X-ray diffraction. Two initial microstructures produced by austenite reversion treatment (ART) were heated at 100 °C/s to 850 °C, 900 °C, or 950 °C and then held isothermally. Rapid heating alone is insufficient for full austenitization, even above the reference $A_{c3}$ temperature determined under slow heating. Full austenitization, defined by bcc fraction ($f_\alpha$) ≤ 1 wt.%, requires short holding, decreasing from about 8 s at 850 °C to about 2 s at 950 °C. The final stage of austenitization is less sensitive to FA temperature than the early holding stage. The initial ART state mainly shifts the starting austenite fraction, whereas both states show comparable kinetic trends at higher FA temperatures.

## 1. Introduction

The infrastructure required for safe energy storage and transportation places strict demands on structural steels, because they must resist failure in hydrogen-containing environments and at low temperatures. In hydrogen energy systems, steels used for pipelines and pressure vessels are exposed to environments where hydrogen ingress can reduce their ductility and fracture toughness [1–3]. This makes hydrogen embrittlement and hydrogen-assisted cracking a central concern for the safe operation and lifetime design of these systems. Meanwhile, cryogenic storage and transport systems, such as those for liquefied natural gas and liquid hydrogen, require base materials and welded joints with sufficient low-temperature toughness, often assessed by impact testing at −196 °C [4,5]. Under these conditions, brittle fracture becomes more likely, and the welded heat-affected zone (HAZ) is often a critical weak region [6]. Taken together, hydrogen and cryogenic service demand steels with a balanced combination of strength, toughness, and weldability, together with microstructures designed to resist hydrogen-induced degradation and low-temperature fracture. In this context, medium manganese steels (MMnSs) have attracted increasing interest as structural materials for energy-related infrastructure, as their multiphase microstructures and austenite stability can be tailored by alloying and heat treatment to improve strength, toughness, and resistance to hydrogen-assisted degradation [7–10]. Their weld HAZ response has also been studied in resistance spot welding and laser welding, where optimized welding schedules and short post-weld thermal cycles have been proposed to control local hardening and improve joint performance [6,11–13].

In MMnSs, the microstructural design strategy centers on a multiphase structure containing austenite with ferrite and/or martensite, allowing strength, ductility, and toughness to be tuned through phase fraction, morphology, and austenite stability. During austenite reversion treatment (ART), Mn partitions into reverted austenite; however, complete Mn homogenization is kinetically limited during short holding because of slow Mn diffusion. Thus, local Mn inhomogeneity can remain after ART. Atom probe tomography (APT) studies have reported core–shell Mn gradients in reverted austenite in MMnSs [14]. while studies on triplex MMnSs have shown local C and Mn segregation at phase boundaries, indicating that nanoscale Mn heterogeneity is strongly processing-dependent [15]. Local Mn enrichment can increase austenite stability and therefore provides a route to tune deformation behavior and the balance between strength, ductility and toughness [16,17]. For hydrogen-related service, such chemical heterogeneity can also be used as a designed feature rather than treated as a defect. Sun et al. [18], for example, introduced a high density of Mn-rich zones, which increased local austenite stability and hydrogen trapping, slowed hydrogen-assisted microcrack growth, and improved hydrogen resistance without sacrificing strength and ductility in high-strength steels. Austenite stability is also important for cryogenic service, where retained austenite stability has been linked to cryogenic impact toughness in a low-C MMnS [19]. Therefore, the mechanical performance of MMnSs for energy applications is closely linked to the heat treatment path and phase transformation kinetics, because they determine the austenite fraction, grain size, austenite stability and Mn heterogeneity.

The concept of flash austenitization (FA) can be traced back to the rapid austenitizing treatments described by Grange [20], in which steel is rapidly heated through the critical temperature range between $A_{c1}$ and $A_{c3}$, with no holding or only a very short holding time, followed by immediate cooling. These cycles are typically produced by induction or resistance heating, with heating rates from ~10 °C/s to several hundred °C/s [21–23]. The austenitization temperature can therefore be reached within seconds, reducing high-temperature exposure during heating and holding. This limited exposure restricts interface migration and diffusion, thereby suppressing austenite grain coarsening [21–23]. Subsequent rapid heating studies have reported reported refined prior austenite grains and finer martensitic substructures after quenching, with stronger refinement at higher heating rates [24–26]. These results support the basic rationale of FA, indicating that austenite formation can occur within a very short high-temperature interval while limiting prior austenite grain coarsening.

For MMnSs, FA is particularly relevant because short high-temperature exposure can promote rapid austenite formation while limiting substitutional Mn diffusion and redistribution. As a result, Mn heterogeneity generated during ART is less likely to be fully homogenized during subsequent austenitization, even after completion of the austenite transformation. Previous studies on MMnSs showed that rapid heating at 100 °C/s enabled fast austenite formation and C partitioning while retaining Mn-related chemical heterogeneity in post-ART or tempered microstructures [27,28]. These findings support a processing route in which ART first introduces Mn segregation and gradients, followed by FA to form austenite rapidly, limit prior austenite grain coarsening, and preserve part of the inherited Mn heterogeneity. However, this route requires transformation kinetics that allow full austenitization within a very short time.

A previous *in situ* synchrotron X-ray diffraction (SYXRD) study showed that heating rate and initial microstructure can be used to control the austenitization path during rapid heating [29]. Increasing the heating rate from 0.25 to 100 °C/s shifted austenitization to higher temperatures and changed the phase-dissolution sequence, including the coupled evolution of ferrite, cementite, and austenite. This is particularly relevant for MMnSs after ART, where ferrite, retained austenite, and martensite coexist before FA and Mn-related chemical heterogeneity may affect local transformation kinetics. However, holding conditions during FA are still often chosen empirically [22,24,25,27], because quantitative evidence on the time scale of seconds required for full austenitization remains limited. In particular, it is still unclear whether full austenitization can be reached without holding or within only a few seconds, and how this depends on austenitization temperature, holding time, and initial ART microstructure. To address this gap, the present work directly quantifies phase evolution during FA after ART using dilatometry-integrated *in situ* SYXRD. FA temperature (850, 900, and 950 °C), holding time, and two initial ART states (ART700-10 and ART700-60) are varied at a fixed heating rate of 100 °C/s to determine the minimum time required for full austenitization and clarify the effects of initial microstructure and holding parameters on transformation kinetics. Based on these results, a transformation-kinetics-based processing window is established to guide FA treatment design for MMnSs.

## 2. Materials and methods

### 2.1 Materials and processing

A MMnS with a nominal composition of Fe-6Mn-1.5Si-1Cr-0.3Mo-0.05Nb-0.2C (wt.%) was produced by vacuum induction melting and cast into a 75 kg ingot with dimensions of ~140 mm × 140 mm × 450 mm. Nb and Mo were added to refine the prior austenite grain size and retard austenite grain growth during subsequent heat treatments [30]. The ingot was homogenized at 1250 °C for 5 h and hot forged between 1200 °C and 900 °C into billets of ~70 mm × 70 mm × 900 mm. The billets were hot rolled in nine passes to 2.0 mm and then cold rolled to 1.5 mm without interpass annealing; no obvious surface defects or edge cracks were observed after cold rolling. Specimens for subsequent experiments were wire-cut from the cold-rolled (CR) sheets, with the 10 mm length parallel to the rolling direction. The specimen geometry for dilatometry and dilatometry-integrated *in situ* SYXRD measurements was 10 mm × 4 mm × 1.5 mm.

The chemical composition was measured by pulse discrimination analysis-optical emission spectrometry (PDA-OES) at Georgsmarienhütte Gruppe and is listed in **Table 1**. The critical transformation temperatures $A_{c1}$, $A_{c3}$, and $M_s$ of the CR condition were determined by dilatometry using a DIL 805 A/D in quenching mode at the Institute of Metal Forming, TU

Bergakademie Freiberg. These temperatures correspond to the start and finish of austenite formation during heating, and the martensite start temperature during cooling under the applied dilatometry conditions. The cycle consisted of heating to 1100 °C at 3 °C/min, holding for 30 min, and quenching to room temperature at 140 °C/s with a fully opened gas valve. **Fig. 1**a shows the corresponding dilatation curves and the transformation temperatures, with $A_{c1}$ = 618 °C, $A_{c3}$ = 776 °C, and $M_s$ = 235 °C. According to the Stahl-Eisen-Prüfblätter (SEP) 1681 standard, the critical temperatures were determined by the lever rule using transformation progress (TF) values of 0.01 and 0.99 as thresholds for transformation start and finish, respectively. Besides the main austenite formation range, the dilatation curve shows a second high-temperature inflection between 865 °C and 1058 °C, which may indicate delayed transformation or dissolution processes in chemical or deformation-heterogeneous regions of the CR material.

Based on the dilatometry results, two ART conditions were applied in a salt bath (GS 540/R2 salt) at 700 °C for 10 min (ART700-10) and 60 min (ART700-60), followed by air cooling to room temperature. These ART conditions served as the initial microstructural states for the subsequent FA experiments. The CR sample and ART-treated microstructures were characterized by electron backscatter diffraction (EBSD) and *ex situ* SYXRD, as described in Section 2.2. *In situ* SYXRD during rapid heating at 100 °C/s and isothermal holding was performed using a DIL 805 A/D at beamline P07B-EH1, as detailed in Section 2.3. Fig. 1b schematically summarizes the ART routes and subsequent in situ FA measurements.

**Table 1.** Chemical composition (in wt.%) of the investigated medium Mn steel.

| MMnS | Fe | Mn | Si | Cr | Mo | Nb | C | P | S | N |
|---|---|---|---|---|---|---|---|---|---|---|
| wt.% | balance | 5.920 | 1.560 | 1.070 | 0.330 | 0.052 | 0.210 | 0.004 | 0.004 | 0.005 |

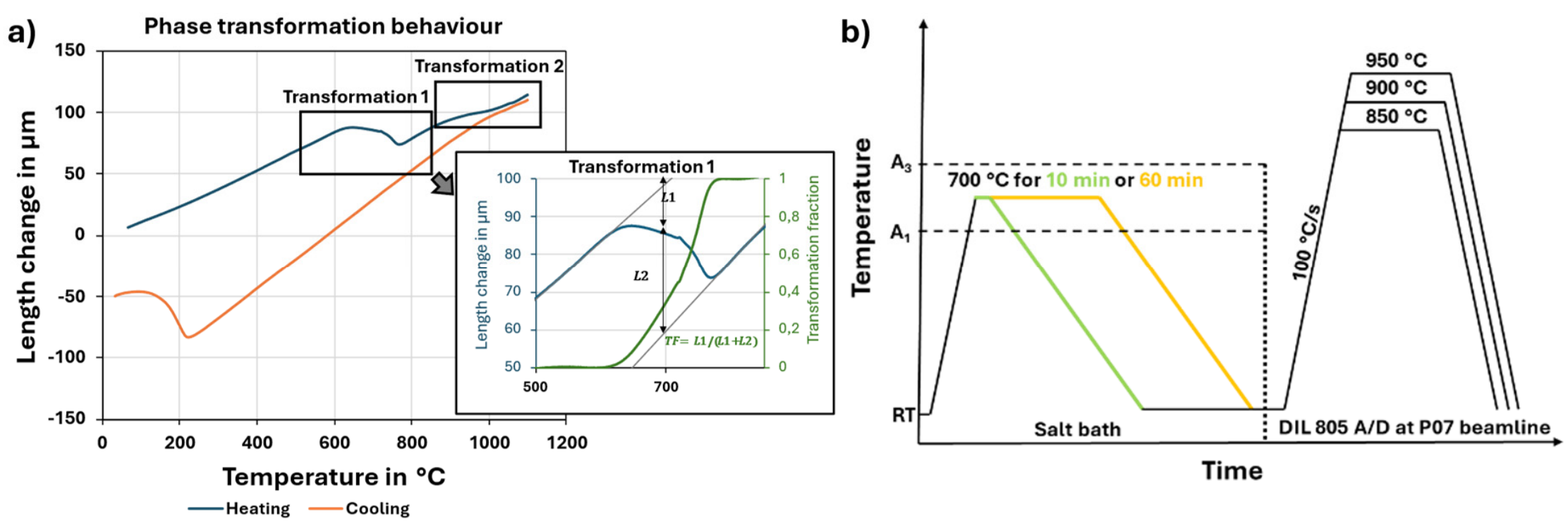

**Fig. 1.** (a) Dilatation curve of the cold-rolled MMnS steel and the local transformation area for $A_{c1}$, $A_{c3}$ and $M_s$ temperature determination using the DIL 805A. (b) Schematic illustration of the austenite reversion treatment (ART) routes at 700 °C for 10 min and 60 min, followed by air cooling, which were used as the initial states for the subsequent *in situ* FA experiments (FA: flash austenitization, RT: room temperature).

## 2.2 Initial microstructure characterization

Microstructure characterization of the CR condition and the two ART conditions (ART700-10 and ART700-60) was carried out by combining EBSD and *ex situ* SYXRD to obtain information on phase constitution, morphology, and crystallographic features at room temperature. EBSD was performed using a Helios 5 Hydra plasma-focused ion beam scanning electron microscope (PFIB-SEM) equipped with a CMOS-based EBSD detector (Oxford Instruments). Specimens were mechanically ground with SiC papers up to 4000 grit, polished with 3 μm and 1 μm diamond suspensions, and finally electropolished in universal electrolyte A2 at 40 V, electrolyte flow-rate setting 18, for 10–12 s. EBSD data were acquired at 20 kV, 13 nA, a 10 mm working distance, and a 0.04 μm step size. Data acquisition and post-processing were conducted using AZtec and AZtecCrystal (Oxford Instruments, Version 3.3), respectively. EBSD maps were used to quantify face-centered cubic (fcc) and body-centered cubic (bcc) phase distributions, grain morphology, and grain size in the CR and ART states. Grain size analysis used a 15° high-angle grain boundary criterion, and grain size was evaluated as the arithmetic mean equivalent circular diameter (ECD). *Ex situ* SYXRD patterns at room temperature were obtained for the CR and ART conditions using the same P07 beamline setup as described for the *in situ* experiments in Section 2.3. The 2D diffraction images were azimuthally integrated over 0° to 360° to obtain 1D intensity profiles using Pydidas [31]. The fcc ($\gamma$) and bcc ($\alpha$) reflections within a $2\theta$ range of 3.5° to 8.5° were identified using crystallographic reference data from the Crystallography Open Database. The diffraction profiles were used for phase identification and for determining the initial peak positions and lattice parameters prior to *in situ* heating.

### 2.3 Dilatometry-integrated *in situ* synchrotron X-ray diffraction

Dilatometry-integrated *In situ* SYXRD was carried out during FA treatment at beamline P07B-EH1 (PETRA III, DESY, Hamburg, Germany) [32]. The experiments were performed in transmission geometry using a monochromatic beam (beam energy about 87.1 keV, wavelength 0.14235 Å) with a beam size of 1 mm × 0.7 mm. This beam size was selected to enable short acquisition times while sampling a sufficiently large gauge volume and maintaining adequate diffraction statistics. FA treatment was conducted using a dilatometer (DIL 805 A/D) from TA Instruments integrated at P07B-EH1, as shown in **Fig. 2**. The chamber had X-ray-transparent windows for the incident and diffracted beams, and the sample was induction-heated while held between quartz push rods. Temperature was monitored and controlled by a thermocouple spot-welded to the specimen surface near the X-ray illuminated volume. The chamber was evacuated and backfilled with Ar to protect the sample and limit oxidation during rapid heating and isothermal holding. Diffraction patterns (Debye–Scherrer rings) were recorded continuously using a PerkinElmer 2D area detector, covering up to $2\theta$ = 8.5°. Detector geometry was calibrated with a $LaB_6$ powder standard (NIST SRM 660C) to determine the beam center, detector tilt and rotation, and sample-to-detector distance.

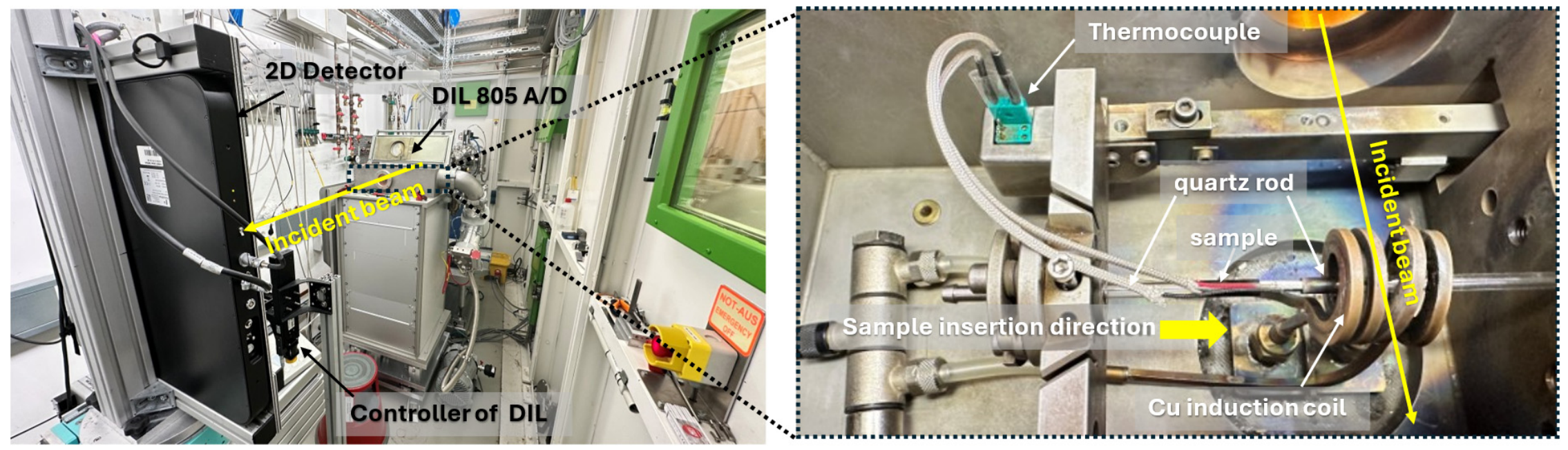

**Fig. 2.** Experimental setup for *in situ* SYXRD during FA at beamline P07B-EH1 (PETRA III, DESY). Left: overview of the DIL 805 A/D mounted in the beamline hutch with the incident beam path and the 2D detector. Right: the chamber showing the specimen positioned between quartz push rods, with the incident beam passing through the gauge section in transmission geometry.

For the *in situ* SYXRD during FA, two initial microstructural states were prepared by ART at 700 °C for 10 min (ART700-10) and 60 min (ART700-60). Specimens from each state were mounted in the DIL 805 A/D and measured throughout the thermal cycle. As shown in Fig. 1b, samples were heated at 100 °C/s to FA temperatures of 850, 900, or 950 °C and held isothermally for 5 min to monitor microstructural evolution and phase transformation. Rapid heating was recorded in fast-acquisition mode at 10 patterns $s^{-1}$, corresponding to a time resolution of 0.1 s, with the beam continuously on. This mode started 30 s before heating, covered the heating stage, and continued for 2 min after reaching the FA temperature. Acquisition was then switched to one pattern every 5 s for the remaining holding period to reduce 2D detector load and data volume. During FA, the DIL 805 A/D recorded the time–temperature profile, whereas the 2D detector stored diffraction data only as sequential image numbers without time stamps referenced to the heat-treatment start. Because the acquisition-mode switch introduced time offsets, the diffraction time axis was reconstructed by synchronizing the dilatometer time–temperature data with the 2D detector sequence using a Python script provided by beamline P07. The script assigned each image number an absolute time and corresponding sample temperature, producing a consistent time–temperature–diffraction dataset for further analysis.

The recorded 2D diffraction images were batch processed by azimuthal integration over 0°–360° to obtain 1D diffraction profiles using Pydidas [31]. Quantitative analysis of the 1D diffraction data was performed by Rietveld refinement using MAUD software (Version 2.99993) [33,34] to obtain the Rietveld-refined phase fractions of fcc ($\gamma$) and bcc ($\alpha$) phases. Phase identification and lattice parameter tracking during heating and holding were performed using the characteristic fcc ($\gamma$) and bcc ($\alpha$) reflections within a $2\theta$ range of 3.5° to 8.5°, with reference crystallographic data taken from the Crystallography Open Database. The instrumental line broadening was determined from a $LaB_6$ standard measured under the same detector geometry. The $LaB_6$ reference pattern was refined to determine the instrumental peak-profile contribution by fitting the background and instrument broadening parameters, including Caglioti, asymmetry and Gaussian parameters in the MAUD instrument model. The resulting instrumental profile parameters were then fixed for refinement of the investigated steel specimens, where only the background, scale factors and lattice parameters of the fcc and bcc phases were refined. The refinement quality was evaluated using the weighted profile $R$ factor ($R_{wp}$), the expected $R$ factor ($R_{exp}$), and $\chi^2$, where $\chi^2 = (R_{wp}/R_{exp})^2$. In MAUD, $\sqrt{\chi^2}$ is represented as *sig*, which corresponds to the goodness of fit ($R_{wp}/R_{exp}$). For correctly estimated experimental uncertainties, $R_{wp}$ approaches $R_{exp}$ and $\chi^2$ approaches 1, whereas $\chi^2 < 1$ may indicate overestimated uncertainties or overparameterization [35]. Since a fixed upper limit for $\chi^2$ is not generally valid and can change with counting statistics, background level, and systematic effects [35], the refinements were also verified by graphical comparison of the observed and calculated patterns, including the difference curve, to confirm that the selected phase and profile models adequately describe the measured data.

## 3. Results

### 3.1 Microstructure characterization of the investigated MMnS in the cold-rolled and ART conditions

The microstructures of the investigated MMnS in the CR condition and after ART at 700 °C for 10 min or 60 min, followed by air cooling to room temperature, were characterized by EBSD and *ex situ* SYXRD. **Fig. 3**a to c show the corresponding EBSD phase maps of the CR, ART700-10 and ART700-60 conditions, respectively, while Fig. 3d shows the corresponding room-temperature *ex situ* SYXRD profiles. The EBSD measurements provide local information on phase distribution, grain morphology, grain boundary misorientation and grain size. The grain size distributions for all studied conditions are provided as supplementary **Fig. S-1**. The quantitative microstructural parameters obtained from EBSD and Rietveld refinement of the SYXRD data are summarized in **Table 2**.

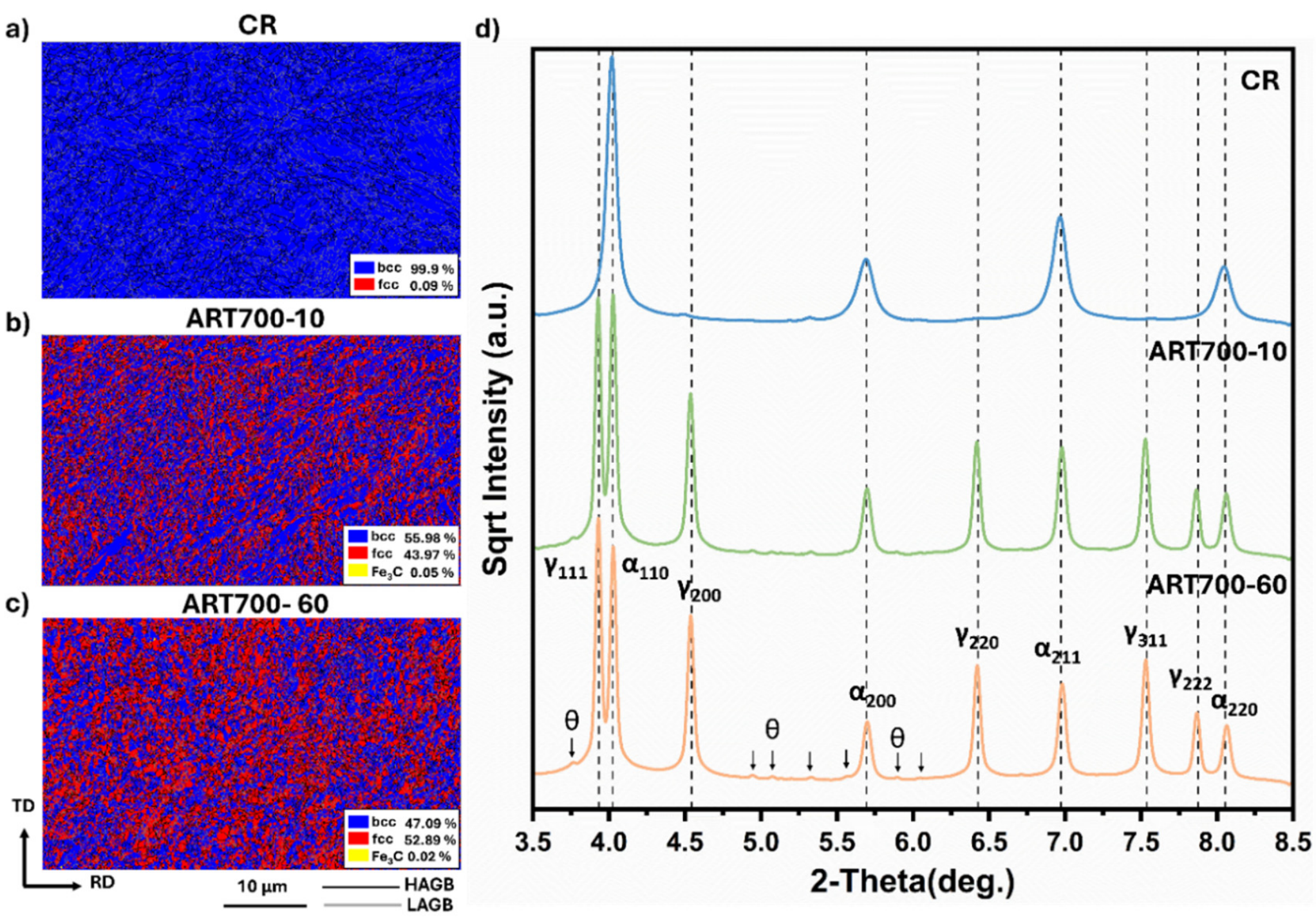

**Fig. 3.** Microstructural analysis of the investigated medium Mn steels (MMnSs) in the cold-rolled (CR) condition and after austenite reversion treatment (ART) at 700 °C. (a) EBSD phase map of the CR condition, (b) EBSD phase map after ART700-10 (700 °C for 10 min, air cooled), and (c) EBSD phase map after ART700-60 (700 °C for 60 min, air cooled). Blue, red and yellow represent the bcc ($\alpha$), fcc ($\gamma$), and cementite ($\theta$) phases, respectively. (d) *Ex situ* SYXRD profiles of the three conditions.

The CR sample (Fig. 3a) shows an almost fully bcc microstructure, with an EBSD-indexed bcc phase fraction of 99.91 %, and an fcc phase fraction of only 0.09 %. Most bcc grains are elongated by rolling, although locally equiaxed grains a few micrometers in size occur between them. After ART at 700 °C for 10 min and air cooling to room temperature (Fig. 3b), the microstructure changes to a refined fcc/bcc dual-phase state, with submicrometer- to micrometer-scale grains. A minor amount of cementite-related indexed regions is detected but should be interpreted cautiously because it is close to the EBSD phase-indexing limitation. The fcc grains are mainly equiaxed and preferentially form along grain and subgrain boundaries of the elongated bcc regions inherited from the CR state, while part of the bcc phase still retains the rolling-induced elongated morphology. Extending ART at 700 °C to 60 min (Fig. 3c) produces a predominantly equiaxed dual-phase microstructure, with slightly coarser fcc grains and a much weaker elongated bcc morphology. Thus, ART transforms the deformation-induced, bcc-dominated CR microstructure into a fine fcc/bcc dual-phase microstructure with a higher austenite fraction and a moderate reduction in low-angle grain boundaries (LAGBs) (Table 2). The EBSD-indexed fcc fraction increases from 43.97% after 10 min to 52.89% after 60 min.

The *ex situ* SYXRD profiles in Fig. 3d confirm the phase constitution identified by EBSD for the CR, ART700-10 and ART700-60 conditions. The dashed vertical reference lines indicate the calculated peak positions of the main fcc and bcc reflections, based on the incident X-ray wavelength and reference lattice parameters of $a_{fcc}$ = 3.591 Å and $a_{bcc}$ = 2.867 Å from the Crystallography Open Database. The calculation procedure is provided as supplementary **Note S-1.** These calculated positions were used for peak assignment, whereas the lattice parameters and phase fractions listed in Table 2 were obtained from Rietveld refinement.

**Table 2.** Microstructural parameters determined from EBSD and *ex situ* SYXRD for the investigated conditions. γ and α denote fcc (austenite) and the bcc (ferrite/martensite-related) phase, respectively. ECD denotes the equivalent circular diameter determined from EBSD. The SYXRD phase fractions are Rietveld-refined weight fractions.

| Sample | EBSD | | | | | | *ex situ* SYXRD | | | |
|---|---|---|---|---|---|---|---|---|---|---|
| | Phase fraction (%) | | Grain size (ECD, μm) | | Grain boundary fraction (%) | | Phase fraction (wt.%) | | Lattice parameter (Å) | |
| | $\gamma$ | $\alpha$ | $\gamma$ | $\alpha$ | LAGB | HAGB | $\gamma$ | $\alpha$ | $\gamma$ | $\alpha$ |
| **CR** | 0.09 | 99.91 | - | 0.68 ± 0.63 | 54.86 | 45.14 | 0.12 | 99.88 | 3.588 | 2.871 |
| **ART700-10** | 43.97 | 55.98 | 0.30 ± 0.18 | 0.42 ± 0.39 | 49.60 | 50.40 | 56.71 | 43.29 | 3.596 | 2.865 |
| **ART700-60** | 52.89 | 47.09 | 0.35 ± 0.23 | 0.44 ± 0.34 | 49.55 | 50.45 | 64.58 | 35.42 | 3.597 | 2.864 |

In the CR condition (blue curve in Fig. 3d), the diffraction profile is dominated by bcc ($\alpha$) reflections (e.g., $\alpha_{110}$, $\alpha_{200}$, $\alpha_{211}$), while no distinct fcc ($\gamma$) reflections are observed. This result is consistent with the EBSD phase map, which shows an almost fully bcc microstructure in the CR state. The bcc ($\alpha$) peaks in the CR conditions are shifted slightly to lower $2\theta$ values compared with the reference positions, corresponding to a larger refined bcc lattice parameter of 2.871 Å (Table 2). This expanded bcc lattice parameter is attributed to the average diffraction signal of deformed ferrite and body-centered tetragonal (bct) martensite phase. Possible contributions to the peak shift and peak broadening may also include C supersaturation in martensite phase, high dislocation density, and residual lattice strain [36–38].

After ART annealing at 700 °C, distinct $\gamma$ reflections ($\gamma_{111}$, $\gamma_{200}$, $\gamma_{220}$, $\gamma_{311}$, $\gamma_{222}$) appear in both ART700-10 and ART700-60, confirming the formation and retention of austenite after air cooling. The refined austenite lattice parameter is 3.596114 Å for ART700-10 and 3.596563 Å for ART700-60, which are rounded to 3.596 Å and 3.597 Å in Table 2, respectively. Both values are slightly larger than the reference fcc lattice parameter, and the value after 60 min ART is only marginally higher. This is consistent with solute redistribution between $\alpha$ and $\gamma$ during ART, especially C enrichment of austenite, because interstitial C expands the fcc lattice [39,40]. However, the small change in the average austenite lattice parameter suggests that extending ART from 10 to 60 min mainly increases the retained austenite fraction rather than strongly changing its average lattice parameter. A possible reason is that C partitioning at 700 °C can proceed relatively rapidly toward local equilibrium, whereas long-range Mn redistribution is much slower. Thus, austenite reversion and growth during longer ART holding are mainly controlled by interface migration rather than local solute redistribution [41–43]. Therefore, the increase in retained austenite fraction with longer ART holding is more pronounced than the change in the average austenite lattice parameter. In addition, weak cementite-related reflections are visible under ART conditions, consistent with the small cementite-related indexed regions observed by EBSD.

Compared with ART700-10, the ART700-60 shows stronger $\gamma$ reflections and weaker $\alpha$ reflections, indicating a higher austenite fraction after prolonged ART holding. This trend is confirmed by Rietveld refinement, with the $\gamma$ fraction increasing from 56.71 ± 0.69 wt.% in ART700-10 to 64.58 ± 0.64 wt.% in ART700-60 (Table 2). The same trend is also observed by EBSD, although the absolute austenite fractions obtained by EBSD are lower than those obtained by SYXRD. This difference is expected because EBSD provides local surface-based area fractions, whereas SYXRD probes a much larger material volume and provides bulk-averaged Rietveld-refined weight fractions. In addition, EBSD can underestimate finely divided or highly strained retained austenite because of step size limits, interaction volume and indexing quality [44,45]. Despite this difference in absolute values, both methods consistently show that extending the ART holding time promotes austenite reversion. Such an increase in retained austenite fraction with increasing ART holding time has been widely reported for MMnSs and is commonly attributed to continued austenite reversion and ongoing elemental partitioning during ART holding [41,46–48].

### 3.2 Dilatometric analysis of transformation temperatures during flash austenitization (FA)

The FA treatment was conducted in a dilatometer to monitor temperature- and time-dependent length changes during rapid heating, isothermal holding, and subsequent cooling. Because fcc and bcc phases have different specific volumes, the dilatation signal provides complementary information on phase transformations throughout the thermal cycle. Transformation temperatures are governed by both the initial microstructure, including chemical constitution, phase fractions, and pre-strained state, and process parameters such as heating rate, which shift the transformation according to material kinetics. Here, the initial conditions were adjusted by the preceding ART, producing two starting states, ART700-10 and ART700-60. Therefore, dilatometry-derived critical transformation temperatures were determined during rapid heating at 100 °C/s for both ART conditions.

**Fig. 4**a shows that both ART-processed conditions exhibit a two-step transformation behavior during rapid heating to 950 °C. For ART700-10, the two transformation stages start at approximately 687 °C and 799 °C, whereas the corresponding temperatures for ART700-60 are approximately 690 °C and 789 °C. Similar transformation temperatures indicate that the start of transformation is only weakly affected by the initial ART state. From the dilatation curves, the finish temperatures of the main austenite formation process are also similar, approximately 875 °C for ART700-10 and 879 °C for ART700-60, indicating that both ART states transform over a comparable temperature range during rapid heating to 950 °C. However, these temperatures should be regarded as dilatometry-derived finish temperatures of bcc-to-fcc transformation rather than direct proof of full austenitization. Therefore, the completion of austenitization during FA needs to be further verified by the *in situ* SYXRD results. The high-temperature transformation segment (transformation 2 as shown in Fig. 4a) is more pronounced in ART700-10 than in ART700-60. This behavior is consistent with the higher bcc phase fraction in ART700-10, which provides a larger phase fraction available for austenite transformation during rapid heating. In contrast, ART700-60 already contains a higher retained austenite fraction before FA, so the additional austenite formation during heating is reduced. These dilatometry results show that the two ART initial states transform at similar temperatures but differ in the amount of transformation occurring during rapid heating.

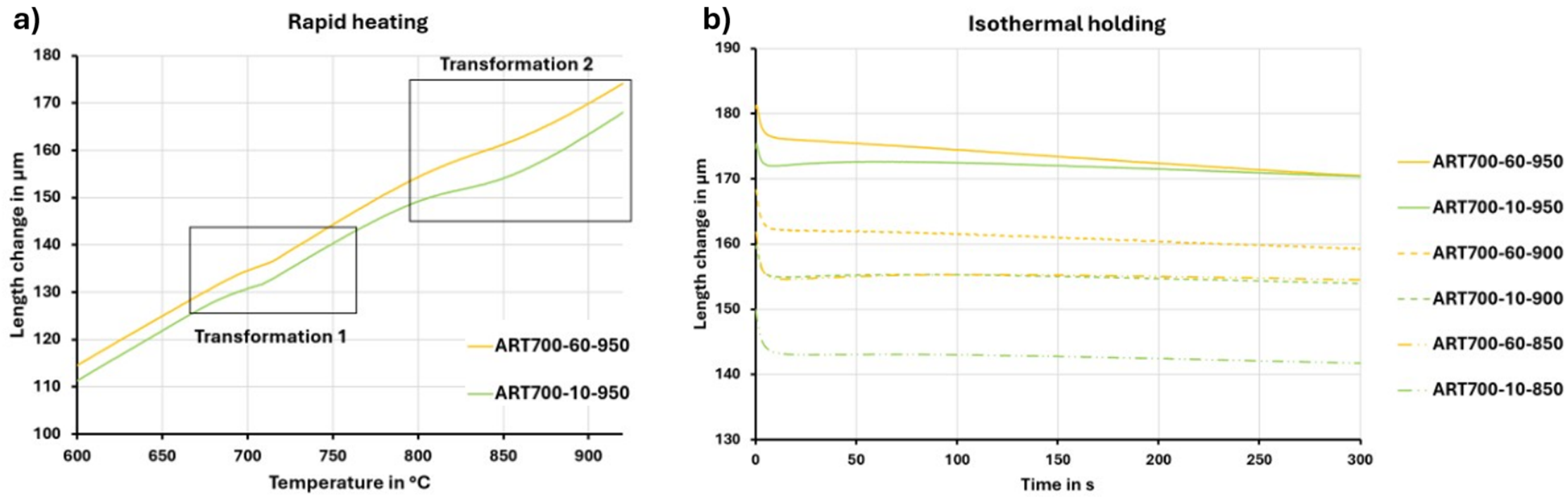

**Fig. 4.** Dilatation behavior of FA during a) rapid heating with 100 °C/s up to 950 °C for both initial microstructural states, ART700-10 and ART700-60, b) dilatation behavior during isothermal holding at the selected austenitization temperatures.

Length reduction during isothermal holding at the FA temperature (Fig. 4b) further indicates incomplete austenitization of the fcc/bcc dual-phase microstructure during rapid heating. At the beginning of the isothermal holding, the length change ($\Delta L$) drops fast and transitions into a more continuous decrease. In some cases, even a small expansion is recognized, which might be caused by a background signal related to the dilatometer setup. This overlaying effect in the $\Delta L$ signal makes it difficult to determine the transformation completion directly from the raw dilatation curve. Therefore, the rate of the relative length change signal was investigated by numerical smoothing using a moving average over eleven consecutive data points and applying equation (1) to the dataset.

$$\frac{\Delta f}{\Delta t} = \frac{f_{i+1} - f_i}{t_{i+1} - t_i} \tag{1}$$

$f$ corresponds to the fraction of length change related to the difference of $\Delta L$ at the beginning and end of isothermal transformation, $i$ is the iteration step and $t$ is the time step.

As shown in **Fig. 5**, the relative isothermal contraction increases rapidly at the start of holding and then either reaches a local maximum followed by a slight decrease or gradually changes into a slower continuous contraction. Larger contraction is observed at lower FA temperatures, indicating that more bcc-to-fcc transformation remains after the target FA temperature is reached. This agrees with the larger untransformed ferrite fraction at the beginning of holding. For ART700-60, higher contraction fractions are observed at 850 and 900 °C than for ART700-10, suggesting more austenite formation during holding. However, because transformation-induced length change cannot be fully separated from overlapping contributions, phase-resolved analysis is required, as discussed in Section 3.3. The corresponding relative contraction rates (Fig. 5b) are high at the beginning of holding and decrease rapidly within the first few seconds. For all conditions, the rate approaches a low, nearly steady level within ~25 s, indicating that the main isothermal transformation is largely completed within this time range.

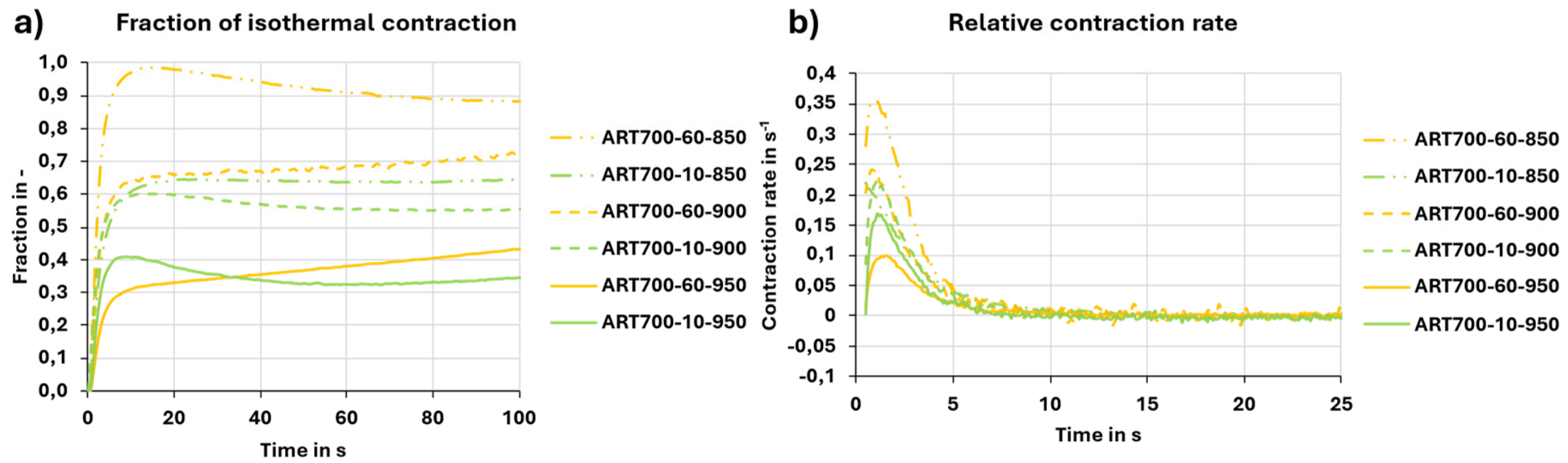

**Fig. 5.** Kinetics of length change during isothermal transformation showing a) the relative fraction of isothermal contraction and b) the corresponding relative contraction rate.

From these dilatation rate data, the end time of the isothermal austenite transformation was estimated and is summarized in **Table 3**. For ART700-60-950, a reliable end time could not be determined due to the smooth transition into continuous contraction. The estimated end time decreases with increasing FA temperature and with longer ART holding before FA. This trend indicates faster transformation kinetics at higher FA temperatures, which is consistent with a higher thermodynamic driving force and faster diffusion at higher temperatures. The shorter end times after ART700-60 may be related to its higher initial austenite fraction and less untransformed ferrite after longer ART, which reduces the amount of transformation required during isothermal holding.

**Table 3.** Estimated end time of the isothermal transformation during FA holding from the contraction rate curves.

| **FA temperature** | **ART700-10** | **ART700-60** |
|---|---|---|
| **950 °C** | 8 s | - |
| **900 °C** | 15 s | 10 s |
| **850 °C** | 22 s | 13 s |

### 3.3 Microstructural evolution and phase transformation behavior during flash austenitization revealed by *in situ* SYXRD

**Fig. 6** summarizes the in situ SYXRD results for ART700-10 during rapid heating in FA at 100 °C/s to 850, 900, and 950 °C. Fig. 6a, c and e present waterfall plots of diffraction profiles over $2\theta$ = 3.5°–6.6°, focusing on temperature ranges near the reference $A_{c1}$ (618 °C) and $A_{c3}$ (776 °C) determined by slow-heating dilatometry in Section 2.1. Fig. 6b, d, f enlarge the $\alpha_{110}$ and $\gamma_{111}$ reflections to more clearly track the bcc ($\alpha$) phase to fcc ($\gamma$) phase transformation during rapid heating. In all three FA thermal cycles, the room temperature diffraction profiles (Fig. 6a, c and e) contain both bcc and fcc reflections, including $\gamma_{111}$, $\gamma_{200}$, $\gamma_{220}$ and $\alpha_{110}$, $\alpha_{200}$, together with weak cementite peaks. This confirms that the three ART700-10 specimens had comparable initial phase constitutions before FA, with the corresponding phase fractions summarized in supplementary **Table S-1**. On heating, the bcc and fcc reflections shift continuously to lower $2\theta$ values, consistent with lattice expansion. Below the reference $A_{c1}$, these shifts mainly reflect thermal expansion and rearrangement of deformed structures, with no substantial bcc-to-fcc transformation. Above $A_{c1}$, the α reflections gradually lose intensity, marking the onset and progress of austenite formation during rapid heating. Near $A_{c3}$, the α intensity drops sharply, indicating transformation of a large fraction of bcc to fcc. Nevertheless, weak $\alpha$ reflections remain after passing $A_{c3}$ and persist until the target FA temperatures are reached, showing that the bcc-to-fcc transformation is incomplete during the rapid heating ramp even at 850, 900, or 950 °C. Cementite-related reflections progressively weaken, indicating dissolution, with only very weak signals remaining at the end of the ramp.

Quantitative phase analysis further confirms that austenitization is incomplete at the end of rapid heating. For the rapid heating to 850 °C (Fig. 6a and b), $\alpha$ reflections remain clearly visible at the end of the heating ramp, especially the $\alpha_{110}$ reflection in the enlarged view. The Rietveld-refined phase fractions at 850 °C are $f_\gamma$ = 90.51 ± 0.54 wt.% and $f_\alpha$ = 9.49 ± 0.39 wt.%, confirming that a considerable bcc fraction remains. A similar trend is observed for rapid heating to 900 °C (Fig. 6c and d). Although the $\alpha_{110}$ intensity is strongly reduced compared with the 850 °C condition, a weak α peak remains detectable at 900 °C, corresponding to $f_\gamma$ = 96.56 ± 1.74 wt.% and $f_\alpha$ = 3.44 ± 0.59 wt.%. Even for rapid heating to 950 °C (Fig. 6e and f), residual $\alpha$ reflections are still detectable at the end of the heating stage, with $f_\gamma$ = 97.62 ± 0.80 wt.% and $f_\alpha$ = 2.38 ± 0.48 wt.%. In this work, full austenitization is defined as $f_\alpha \leq 1$ wt.% (i.e., $f_\gamma \geq 99$ wt.%). This threshold is used as a conservative operational criterion, consistent with transformation analyses in which the start and finish of a transformation are often defined using 1 % and 99 % transformed fractions, respectively [49]. It also avoids overinterpreting very small Rietveld-refined phase fractions close to the typical detection limit of quantitative laboratory X-ray diffraction [50], while providing a consistent basis for comparing all FA conditions. Although synchrotron diffraction can detect weaker phase contributions, the 1 wt.% threshold provides a consistent criterion for comparing the present FA conditions. Using this criterion, none of the three FA treatments reaches full austenitization during rapid heating alone, even at 950 °C.

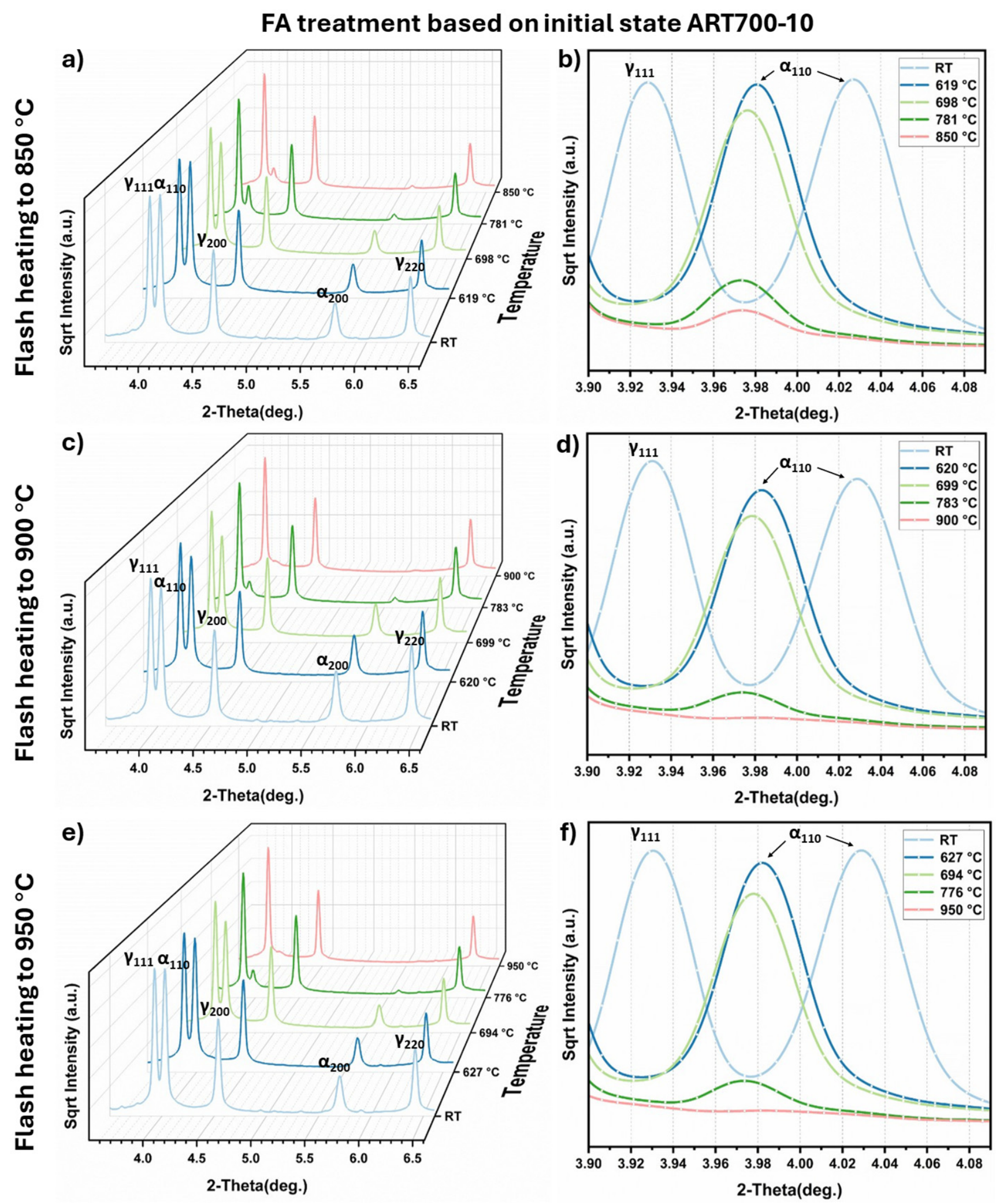

**Fig. 6.** *In situ* SYXRD results show microstructural evolution and phase transformation of ART700-10 during rapid heating at 100 °C/s from room temperature (RT) to FA temperatures of 850 °C (a, b), 900 °C (c, d), and 950 °C (e, f). Fig. 6a, c and e present waterfall plots of the diffraction profiles in the $2\theta$ range 3.5 to 6.6°, highlighting the evolution of bcc ($\alpha$) and fcc ($\gamma$) reflections during heating. Fig. 6b, d, and f show enlarged views around $\gamma_{111}$ and $\alpha_{110}$.

**Fig. 7** shows the *in situ* SYXRD response of ART700-10 during isothermal holding after rapid heating in FA, with time (t) = 0 defined as the moment the target FA temperature is reached. The waterfall plots in Fig. 7a, c, e over $2\theta$ = 3.5° to 6.6° track the diffraction-profile evolution during 300 s holding, while Fig. 7b, d, f enlarge the $\alpha_{110}$ region to monitor residual bcc. Selected phase fractions are summarized in Table S-1. The lower initial $\alpha_{110}$ intensity at 900 and 950 °C than at 850 °C agrees with the smaller $\alpha$ fraction remaining after rapid heating, as shown in Fig. 6 and Table S-1. At 850 °C (Fig. 7a, b), the $\alpha_{110}$ intensity decreases strongly within the first few seconds. The α fraction drops from 9.49 ± 0.39 wt.% at t = 0 to 0.85 ± 0.13 wt.% after 8 s, while $f_\gamma$ increases to 99.15 ± 1.97 wt.%, satisfying the full austenitization criterion of $f\alpha \leq 1$ wt.%. At 900 and 950 °C (Fig. 7c–f), the same trend occurs, but the time required to reach the completion criterion is shorter. At 900 °C, $f_\alpha$ decreases from 3.44 ± 0.59 wt.% to 0.87 ± 0.17 wt.% after 4 s, with $f_\gamma$ = 99.12 ± 0.69 wt.%. At 950 °C, $f_\alpha$ reaches 0.78 ± 0.15 wt.% after only 2 s, giving the shortest completion time among the three FA temperatures.

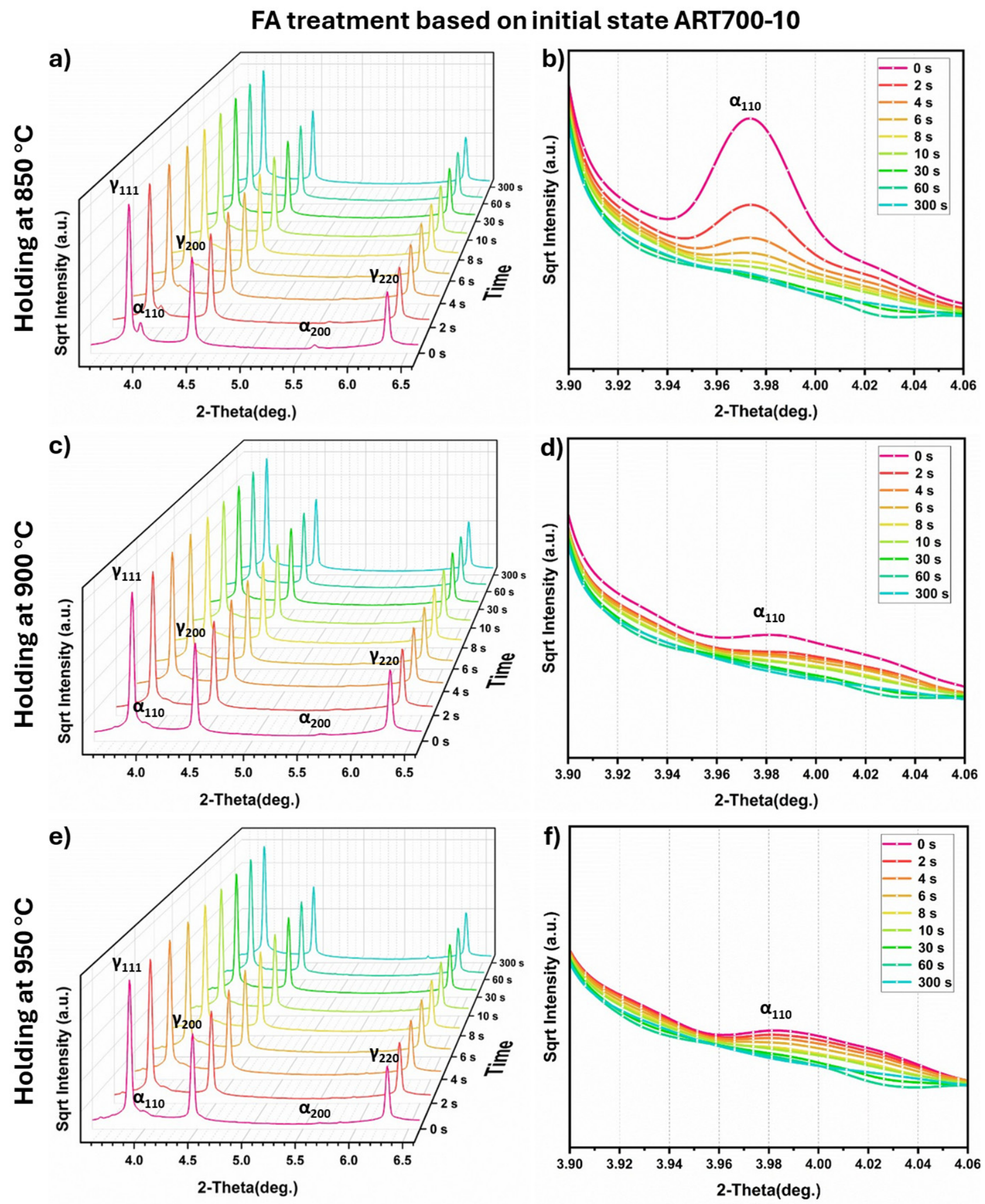

**Fig. 7.** *In situ* SYXRD results of ART700-10 during isothermal holding after rapid heating in FA to 850 °C (a, b), 900 °C (c, d), and 950 °C (e, f). Waterfall plots (a, c, e) show the time-dependent evolution of $\alpha$ and $\gamma$ reflections during holding, and the enlarged views (b, d, f) highlight the progressive decrease of the residual $\alpha_{110}$ reflections.

In addition to the intensity changes, the position of the residual $\alpha_{110}$ reflection also changes during isothermal holding. When comparing the 850 °C, 900 °C and 950 °C conditions, the residual $\alpha_{110}$ peak is located slightly towards higher 2θ at higher FA temperatures, although thermal expansion alone would shift diffraction peaks towards lower $2\theta$. This indicates that the lattice parameter of the small residual $\alpha$ fraction is affected not only by temperature, but also by changes in local composition and strain state during holding. One possible contribution is C depletion of the residual $\alpha$ phase, because interstitial C expands the bcc/bct lattice and its redistribution into $\gamma$ would reduce the $\alpha$ lattice parameter [51]. However, this interpretation should be treated with caution because the residual $\alpha$ fraction is very small at high FA temperatures, and the $\alpha_{110}$ peak partially overlaps with the tail of the $\gamma_{111}$ reflection. Time- and temperature-dependent C and Mn partitioning between $\alpha$ and $\gamma$ is expected during austenite reversion in MMnSs [52], and faster solute redistribution at higher temperature is consistent with the shorter holding times required to reach the full austenitization criterion. During holding at 950 °C, an additional weak peak near $2\theta \approx 5.84°$ gradually appears and increases in intensity. This peak cannot be readily assigned to the main $\alpha$ or $\gamma$ reflections and may indicate the formation of a minor high-temperature secondary phase. A tentative comparison with crystallographic database entries suggests that this feature could be related to a $Cr_3Si$-type structure, such as $(Cr,Mn,Fe)_3Si$.

**Fig. 8** shows the *in situ* SYXRD results for ART700-60 during rapid heating in FA at 100 °C/s to 850, 900, and 950 °C. The room-temperature profiles contain both $\alpha$ and $\gamma$ reflections, confirming a duplex initial microstructure in all three FA cycles, with comparable initial phase fractions listed in Supplementary **Table S-2**. During rapid heating, the $\alpha$ and $\gamma$ reflections shift to lower $2\theta$ values due to thermal expansion, while the α peak intensity decreases progressively, especially near the reference $A_{c3}$ temperature. This indicates rapid bcc-to-fcc transformation during heating, as in ART700-10.

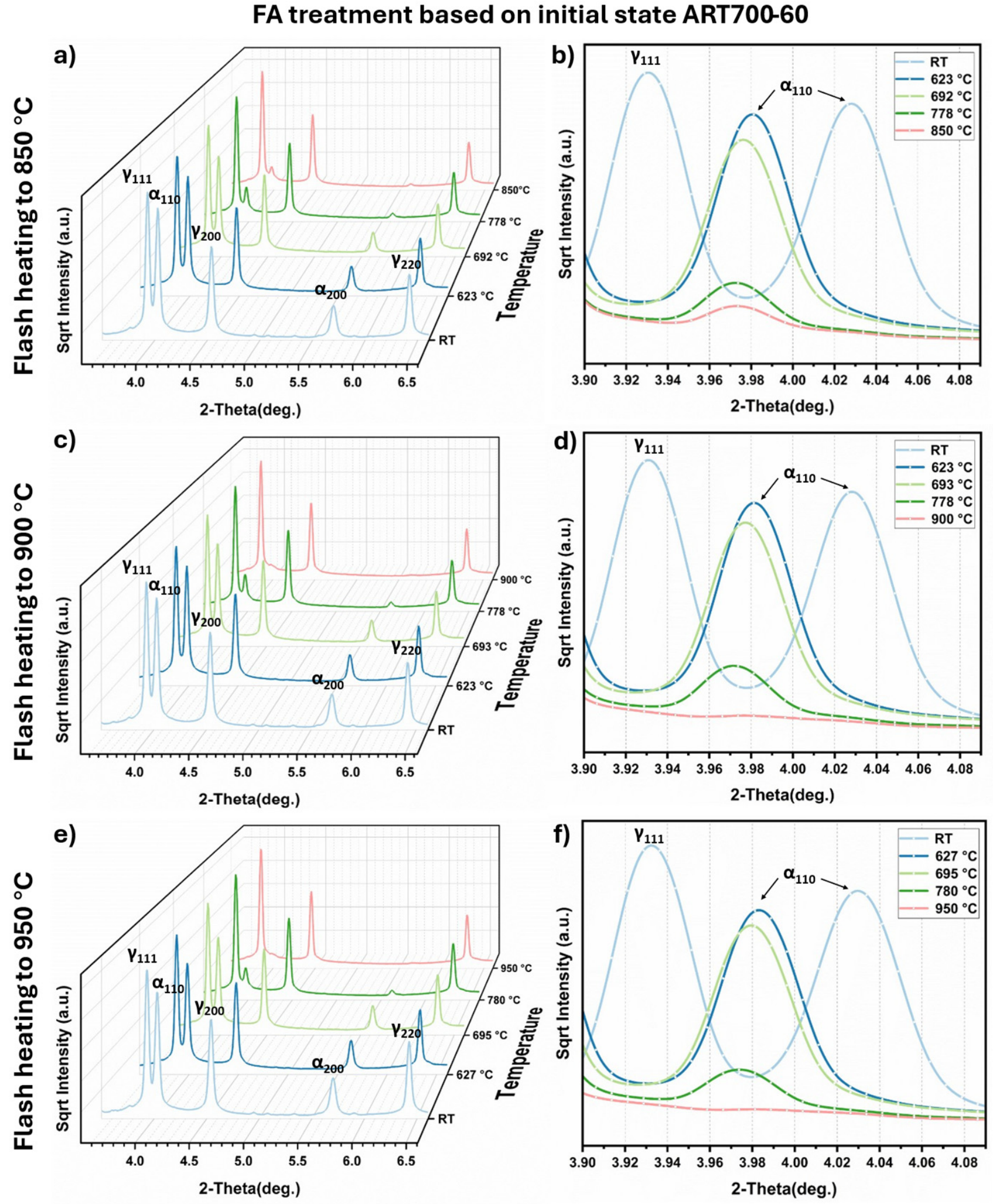

**Fig. 8.** *In situ* SYXRD results show microstructural evolution and phase transformation of ART700-60 during rapid heating in FA at 100 °C/s from room temperature (RT) to FA temperatures of 850 °C (a, b), 900 °C (c, d), and 950 °C (e, f). Fig. 8a, c, e present waterfall plots of the diffraction profiles in the $2\theta$ range 3.5 to 6.6°. Fig. 8b, d, and f show enlarged views of $\gamma_{111}$ and $\alpha_{110}$, highlighting the progressive decrease in $\alpha_{110}$ intensity during rapid heating.

However, residual $\alpha$ remains detectable at the end of the heating ramp for all target temperatures, as shown by the enlarged $\alpha_{110}$ regions in Fig. 8b, d, f. The Rietveld-refined phase fractions at the end of rapid heating confirm incomplete austenitization. At 850, 900, and 950 °C, the γ fractions are $f_\gamma = 92.33 \pm 0.61$, $96.31 \pm 0.27$, and $97.97 \pm 1.47$ wt.%, while the corresponding α fractions are $f_\alpha = 7.67 \pm 0.38$, $3.68 \pm 0.39$, and $2.03 \pm 0.29$ wt.%, respectively. Based on the full austenitization criterion of $f_\alpha \leq 1$ wt.%, none of the FA treatments from the ART700-60 initial state achieve full austenitization during rapid heating alone over the 850–950 °C range. Thus, exceeding the reference $A_{c3}$ determined under slow-heating conditions is insufficient at 100 °C/s without holding.

**Fig. 9** presents *in situ* SYXRD results for ART700-60 during isothermal holding after rapid heating to 850, 900, and 950 °C, with t = 0 defined as the moment the target FA temperature is reached. Selected phase fractions are listed in Table S-2.

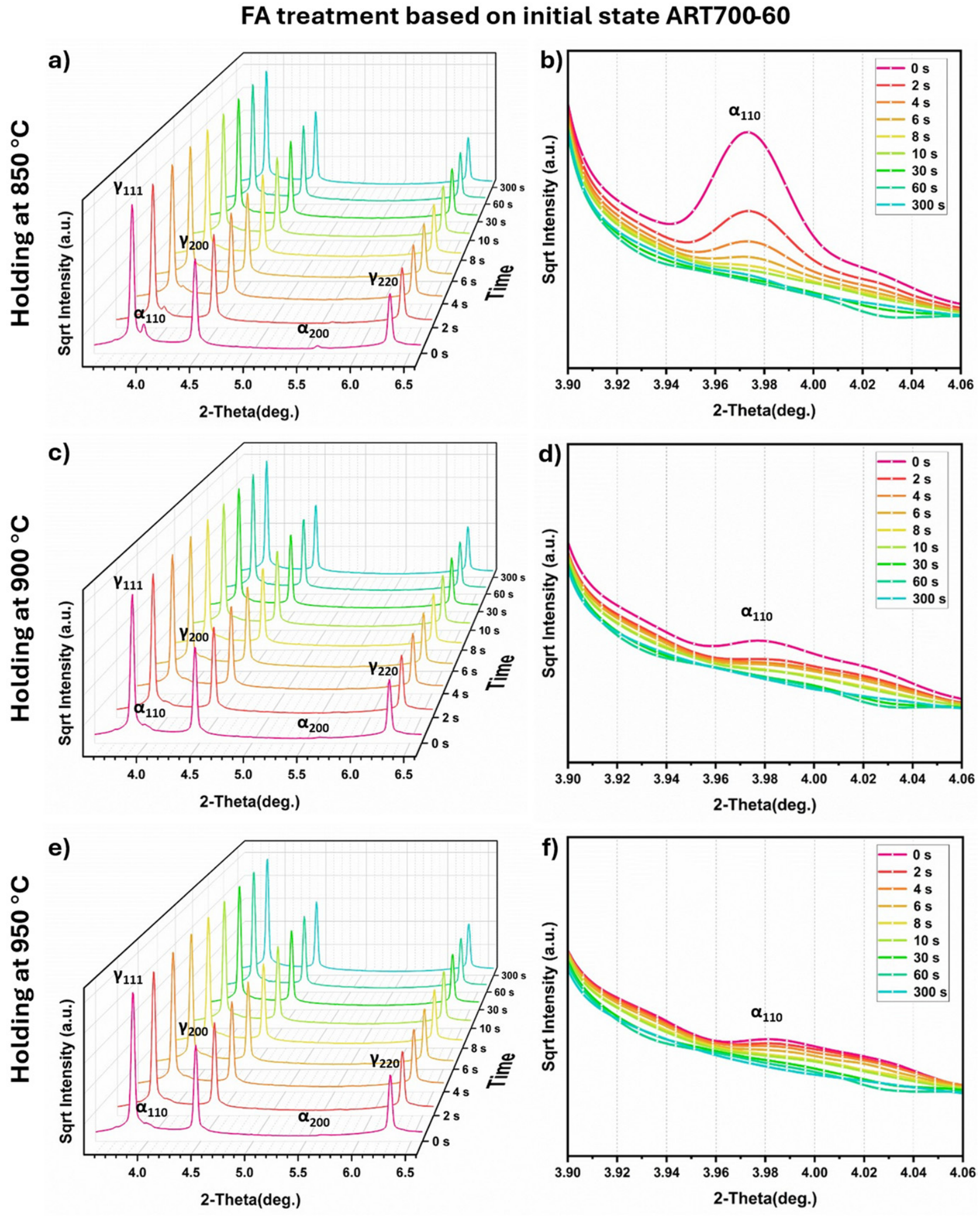

**Fig. 9.** *In situ* SYXRD results show microstructural evolution and phase transformation of ART700-60 during isothermal holding after rapid heating in FA to 850 °C (a, b), 900 °C (c, d), and 950 °C (e, f). Fig. 9a, c, e present waterfall plots of the diffraction profiles in the $2\theta$ range 3.5 to 6.6°, showing the time-dependent evolution of bcc ($\alpha$) and fcc ($\gamma$) reflections during holding. Fig. 9b, d, and f show enlarged views around the $\alpha_{110}$ reflection.

As in ART700-10, residual $\alpha$ remaining after rapid heating decreases rapidly within the first seconds of holding, followed by slower changes once its fraction becomes small. At 850 °C, $f_\alpha$ decreases from 7.67 ± 0.38 wt.% at t = 0 to 0.97 ± 0.14 wt.% after 8 s, while $f_\gamma$ increases to 99.03 ± 1.61 wt.%, satisfying the full austenitization criterion of $f_\alpha \leq 1$ wt.% after ~8 s. At 900 °C, $f_\alpha$ decreases from 3.68 ± 0.39 wt.% to 1.01 ± 0.15 wt.% after 4 s, within the uncertainty range of the criterion. At 950 °C, $f_\alpha$ decreases from 2.03 ± 0.29 wt.% to 1.08 ± 0.17 wt.% after 2 s. Taken together, the *in situ* SYXRD results show that rapid heating at 100 °C/s strongly promotes austenite formation but does not complete the bcc-to-fcc transformation in either ART initial state. Full austenitization is achieved only after a short isothermal holding step, with the required time decreasing as the FA temperature increases.

## 4. Discussion

### 4.1 Coupled effects of FA temperature and holding time on austenitization kinetics

**Fig. 10** summarizes the austenite formation kinetics during FA for two initial ART states, namely ART700-10 (Fig. 10a) and ART700-60 (Fig. 10b). For the austenite fraction curves, blue, green, and purple denote FA to 850 °C, 900 °C, and 950 °C, respectively. Fig. 10c further compares the isothermal holding stage after time normalization, where t = 0 s denotes the start of holding at the target FA temperature.

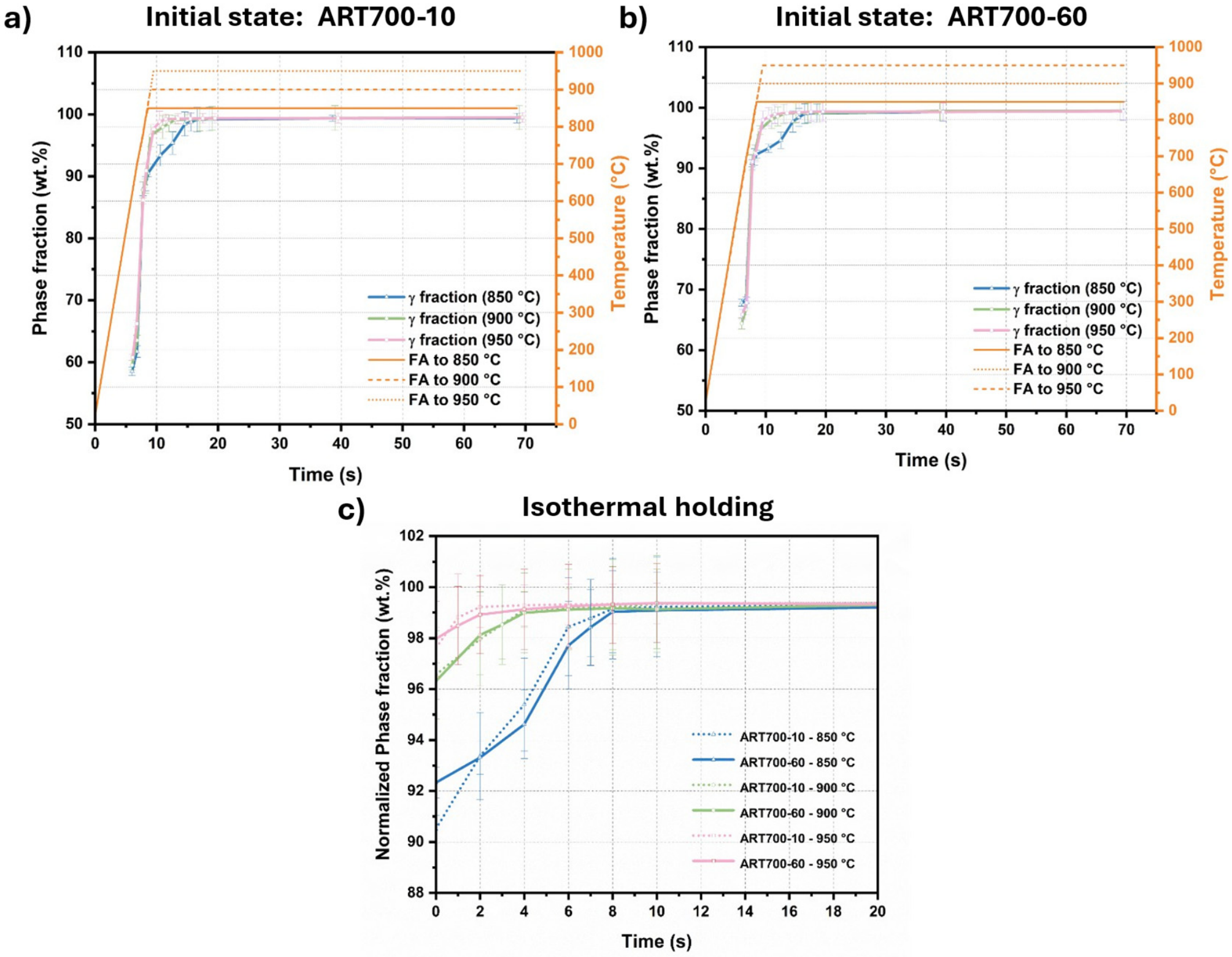

**Fig. 10.** Austenite ($\gamma$) transformation kinetics during FA at different target temperatures of 850 °C, 900 °C, and 950 °C for (a) ART700-10 and (b) ART700-60. The $\gamma$ phase fractions were obtained by Rietveld refinement of the *in situ* SYXRD patterns, while the orange curves show the corresponding temperature-time profiles. (c) Comparison of the $\gamma$ transformation kinetics during the isothermal holding stage after time normalization, where t = 0 s denotes the start of holding at the target FA temperature.

The comparison in Fig. 10 shows that FA temperature affects austenite transformation kinetics in two coupled ways: it enhances transformation kinetics by increasing diffusivity and interface mobility during austenite formation, while also extending the high-temperature segment of the rapid-heating path before isothermal holding. At 100 °C/s, heating from 850 to 950 °C adds 1 s in the temperature range where bcc-to-fcc transformation remains active. This is not negligible, since dilatometry shows that the main transformation during rapid heating is incomplete at 850 °C. Moreover, the austenite fraction at the start of holding at 950 °C is about 5–6 wt.% higher than that after 1 s holding at 850 °C, indicating that transformation at 850 °C remains far from the full-austenitization criterion within the first second of holding. Thus, the higher initial austenite fraction at 950 °C results not only from the higher FA temperature, but also from additional transformation during the final stage of rapid heating. Consequently, early holding kinetics are highly sensitive to FA temperature, as each target temperature yields a different transformation state at the start of holding.

The average reduction rate of the residual α fraction during holding is estimated as $\Delta f_{\alpha}/\Delta t$. If the shorter holding time at higher FA temperature were primarily caused by an acceleration of the isothermal transformation stage, the average reduction rate of residual α would be expected to increase with temperature. Instead, for ART700-10, the rates are approximately 1.10, 0.65, and 0.80 wt.% $s^{-1}$ at 850, 900, and 950 °C, respectively; for ART700-60, the corresponding values are approximately 0.84, 0.68, and 0.48 wt.% $s^{-1}$. Thus, the reduced holding time at higher FA temperature mainly results from the smaller residual $\alpha$ fraction inherited from the heating ramp, rather than from a faster average transformation rate during holding. The holding stage therefore mostly captures the late stage of austenitization close to the full-austenitization criterion. When the residual $\alpha$ fraction is already small, the time required to reduce the remaining $\alpha$ fraction from approximately 1.5 wt.% to approximately 0.8 wt.% is close to 2–3 s for the different FA temperatures. This indicates that the final stage to full austenitization is less sensitive to FA temperature than the early stage of isothermal holding.

The time-normalized holding curves in Fig. 10c further clarify the transformation mechanism. At 850 °C, the curve retains a visible S-shaped character, indicating that the isothermal stage still includes an acceleration period before the transformation rate decreases near completion. This behavior is consistent with a nucleation-and-growth type transformation [53,54], although the transformation already starts during non-isothermal heating and the initial fraction at the start of holding differs between FA temperatures. In contrast, the curves at 900 and 950 °C approach the plateau more directly, indicating that much of the early-stage transformation has already occurred before the isothermal hold begins. The subsequent holding stage is therefore dominated by the consumption of a small amount of residual α and the final approach to full austenitization.

These results show that optimization of FA treatment may not be based on the nominal holding time alone. A higher peak temperature can reduce the required holding time by advancing austenitization during rapid heating, but it also increases the effective high-temperature exposure, which may promote austenite grain coarsening and substitutional solute homogenization [55,56]. The processing question is therefore not only whether full austenitization can be achieved, but also how it can be achieved with the minimum effective high-temperature exposure. Within the investigated 850–950 °C window, a short isothermal holding step provides a controlled route to full austenitization while limiting unnecessary thermal exposure.

Moreover, the weak additional diffraction feature observed during holding at 950 °C may indicate the formation of a minor secondary phase. A possible assignment is a $Cr_3Si$-type transition-metal silicide promoted by high-temperature redistribution of Cr, Mn, and Si [57]. However, this interpretation remains tentative because the present diffraction data alone is insufficient for unambiguous phase identification.

### 4.2 Effect of initial ART microstructure on phase transformation kinetics during flash austenitization

**Fig. 11** addresses whether the different initial ART states modify the austenitization kinetics during FA or mainly shift the initial austenite fraction before rapid transformation proceeds. The $A_{c1}$ and $A_{c3}$ temperatures determined by slow-heating dilatometry, 618 °C and 776 °C, respectively, are used here only as reference markers. Under slow heating, the material has sufficient time for C redistribution and limited local Mn redistribution before the main transformation is detected macroscopically. This can partly reduce the chemical heterogeneity, so that the dilatometry response shows one apparent $A_{c1}$. During FA after ART, in contrast, the heating time is much shorter, so the inherited chemical heterogeneity is partly retained. Regions enriched in C and Mn are expected to be stabilized toward austenite and can transform as austenite at lower temperatures, whereas chemically leaner $\alpha$ regions require higher temperatures and/or additional time to transform. This explains why the FA dilatometry curves show a clearer multi-stage transformation and why the effective transformation temperatures shift to higher values under rapid heating.

The correspondence between dilatometry and SYXRD should be interpreted qualitatively. Dilatometry records a macroscopic length-change response integrated over the specimen, whereas SYXRD provides phase-resolved structural information from the irradiated volume. However, both methods reveal a multi-stage transformation during rapid heating after ART. Near the reference $A_{c1}$ temperature, which corresponds to the first phase-fraction data point in each curve in Fig. 11a, ART700-60 shows a higher austenite fraction than ART700-10. This difference is mainly inherited from the higher retained austenite fraction before FA, rather than from faster newly formed $\gamma$ during the early heating stage. Between the reference $A_{c1}$ and $A_{c3}$ temperatures, the austenite fraction increases rapidly for all FA conditions. For a given FA temperature, the ART700-10 and ART700-60 curves in Fig. 11a show broadly comparable slopes within the refinement uncertainty, and the main difference remains the offset in absolute austenite fraction. This suggests that, once rapid austenite formation starts in this temperature range, the transformation is mainly controlled by temperature-dependent interface migration and solute redistribution, while the initial ART condition mainly determines the starting phase fraction and the retained chemical heterogeneity.

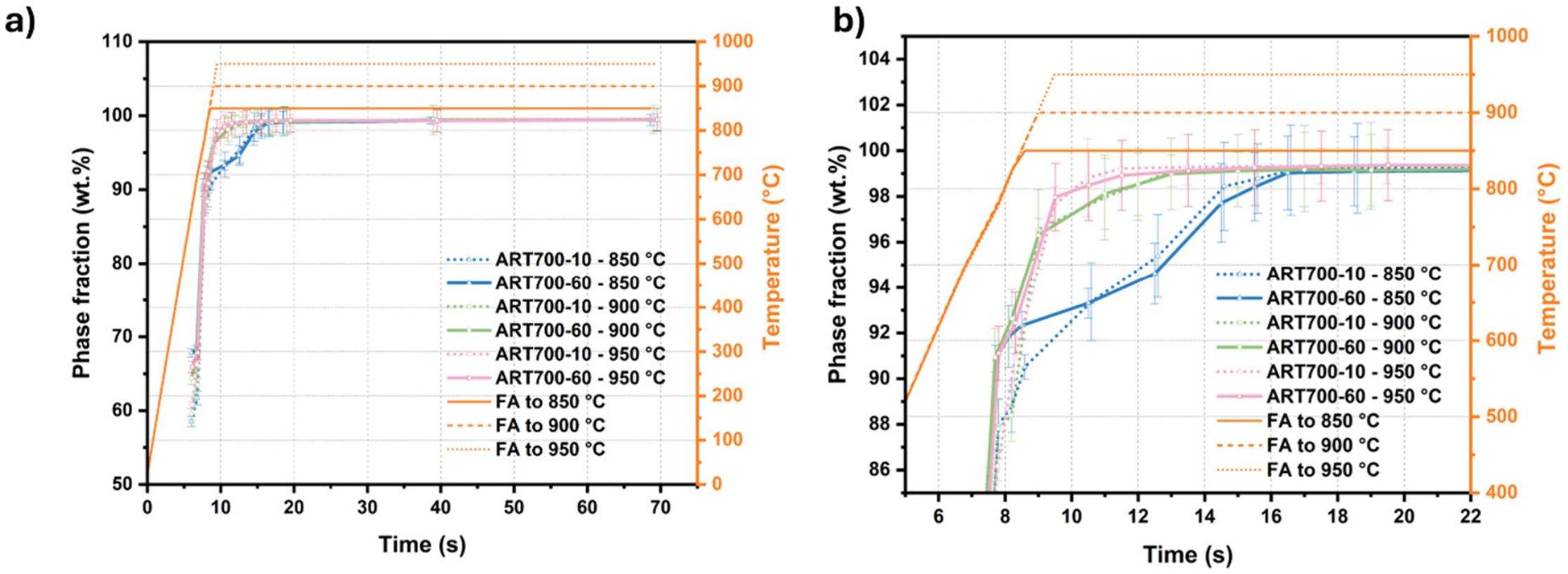

**Fig. 11.** Comparison of austenite (γ) formation kinetics during flash austenitization (FA) for two initial ART states at FA temperatures of 850 °C, 900 °C, and 950 °C. (a) *γ* phase fraction as a function of time for ART700-10 and ART700-60. Dash-dotted curves represent ART700-10, solid curves represent ART700-60, and the orange curves show the corresponding temperature-time profiles. (b) Enlarged view of the late heating and early holding stage, highlighting the kinetic differences between the two initial ART states.

At the reference $A_{c3}$ temperature (776 °C), none of the samples has reached the full austenitization criterion, as shown in Fig. 11b. This indicates that the effective completion temperature under rapid heating is higher than the $A_{c3}$ value determined under slow-heating conditions. This difference is consistent with the known heating-rate dependence of the critical transformation temperature and with the limited time available for interface migration, solute redistribution, and carbide dissolution during rapid heating [29,58–60]. After passing the reference $A_{c3}$ temperature and before reaching the final FA temperature, the austenite formation rate decreases for all conditions. This indicates that the remaining transformation is no longer dominated by rapid early-stage nucleation and growth, but by the final consumption of residual α regions and further local solute redistribution processes [61,62].

Once isothermal holding begins, the influence of the initial ART state on the transformation kinetics becomes less pronounced. For a given FA temperature, ART700-10 and ART700-60 approach a near plateau within comparable kinetics, especially at 900 °C and 950 °C. This convergence indicates that high-temperature holding weakens the influence of the initial ART microstructure by promoting interface migration, C and Mn redistribute between phases [63,64], and continued carbide dissolution [58]. In contrast, at 850 °C, the difference between the two ART states remains more visible because a larger residual α fraction is still present at the start of holding and the subsequent transformation proceeds more slowly. The inherited phase-fraction offset is therefore reduced more slowly at 850 °C than at higher FA temperatures.

Overall, within the investigated ART and FA conditions, the initial ART state mainly affects the austenite fraction at the beginning of FA and during early heating. In contrast, the target FA temperature and short holding time control how rapidly the residual *α* fraction approaches the full-austenitization criterion. At higher FA temperatures, the austenite fraction evolution becomes increasingly similar for ART700-10 and ART700-60 despite their different initial phase fractions. This indicates that the FA processing window for full austenitization is governed primarily by the imposed thermal path, while the initial ART microstructure shifts the starting point of the transformation through the inherited austenite fraction and retained local chemical heterogeneity.

The dilatometry-derived contraction-rate analysis supports the same temperature dependence as the *in situ* SYXRD results, but it gives longer end times for full austenitization during the isothermal holding stage. For ART700-10, the end times estimated from the dilatation-rate curves are 22 s, 15 s and 8 s at 850 °C, 900 °C and 950 °C, respectively, whereas the SYXRD-based full austenitization criterion is reached after about 8 s, 4 s and 2 s. For ART700-60, the same discrepancy is observed. This systematic difference is expected because dilatometry records the macroscopic length change of the specimen, which can include transformation-related anisotropic contraction [65], overlapping microstructural evolution [66], as well as instrumental contributions associated with longitudinal temperature gradients due to the rapid induction heating [67]. In contrast, a simultaneous SYXRD study can directly track the evolution of individual phase fractions during austenitization [29]. Therefore, the two methods are consistent in trend, but *in situ* SYXRD provides the more direct basis for defining the full austenitization criterion and the FA processing window.

## 5. Conclusions

This study quantified austenite formation during FA of a MMnS after ART using dilatometry-integrated *in situ* SYXRD. FA was conducted at 100 °C/s to 850 °C, 900 °C, or 950 °C, followed by isothermal holding. Two initial ART states, ART700-10 and ART700-60, were compared. Full austenitization was evaluated using the criterion of $f_\alpha \leq 1$ wt.%.

The main conclusions are as follows:

- Rapid heating to a FA temperature above reference $A_{c3}$ determined under slow-heating conditions is not sufficient for full austenitization. For both initial ART states, a detectable residual $\alpha$ remains at the end of rapid heating, even at 950 °C.
- FA temperature affects austenitization kinetics by changing both the driving force for austenite formation and the residual α fraction present at the start of holding. Increasing the FA temperature from 850 °C to 950 °C shortens the holding time required for full austenitization. The final stage of austenitization is less sensitive to FA temperature than the early holding stage.
- The initial ART state mainly shifts the starting austenite fraction. ART700-60 has a higher initial $\gamma$ fraction and retains this offset during early rapid heating, but both ART states show comparable kinetic slopes through the reference intercritical range within uncertainty.
- The influence of the initial ART state becomes less pronounced with increasing FA temperature and holding time. It remains more visible at 850 °C and during the first seconds of holding, whereas the $\gamma$ fraction evolution becomes more similar at 900 °C and 950 °C.

Overall, full austenitization during FA requires a matched FA temperature and holding time rather than a peak-temperature criterion alone. Within the investigated range, short holding at 850–950 °C provides a controlled route to complete austenitization while limiting high-temperature exposure. This is important because simply increasing the peak temperature may reduce the required holding time but can also weaken the FA advantage in limiting austenite grain coarsening and substitutional solute homogenization. The weak additional diffraction feature observed during holding at 950 °C for ART700-10 suggests that high FA temperatures may also introduce minor high-temperature second phases, which require further verification. Future work will link this FA processing window to the resulting microstructure and service-related mechanical properties to enable the application of MMnSs in energy-related infrastructure.

**Conflict of Interest**

The authors declare that they have no conflict of interest.

**Data Availability Statement**

The data that support the findings of this study are available from the corresponding author upon reasonable request.

**Acknowledgements**

The authors gratefully acknowledge Deutsches Elektronen-Synchrotron (DESY, Hamburg, Germany) and the staff of beamline P07 at PETRA III, with special thanks to Dr. Emad Mawaad for assistance during the X-ray diffraction experiments. Financial support from the Deutsche Forschungsgemeinschaft (DFG, German Research Foundation) under grant no. 490856143 is gratefully acknowledged.